Jelena Cupać
WZB Berlin Social Science Center
https://orcid.org/0000-0002-7471-7624
jelena.cupac@wzb.eu

Jelena Cupać is a guest Research Fellow at the Global Governance Unit of the WZB Berlin Social Science Center.

# The Gender Code:
# Gendering the Global Governance of Artificial Intelligence

**Abstract**

This paper examines how international AI governance frameworks address gender issues and gender-based harms. The analysis covers binding regulations, such as the EU AI Act; soft law instruments, like the UNESCO Recommendations on AI Ethics; and global initiatives, such as the Global Partnership on AI (GPAI). These instruments reveal emerging trends, including the integration of gender concerns into broader human rights frameworks, a shift toward explicit gender-related provisions, and a growing emphasis on inclusivity and diversity. Yet, some critical gaps persist, including inconsistent treatment of gender across governance documents, limited engagement with intersectionality, and a lack of robust enforcement mechanisms. However, this paper argues that effective AI governance must be intersectional, enforceable, and inclusive. This is key to moving beyond tokenism toward meaningful equity and preventing reinforcement of existing inequalities. The study contributes to ethical AI debates by highlighting the importance of gender-sensitive governance in building a just technological future.

**Introduction**

For all the immense promise that artificial intelligence (AI) holds for social, economic, and technological progress, a reckoning is underway over the myriad risks this new technology poses to marginalized groups, labor, the environment, democracy, and the international order, among others (Bradford, 2023; Crawford, 2021; Cupać et al., 2024). While AI harm in each of these areas is likely to impact women and affect their rights, current debates and academic research have particularly focused on the damage caused by algorithms that display *gender bias*—i.e., unfair differences in treatment or outcomes resulting from algorithms that treat disadvantaged, minority, and marginalized gender groups less favorably than the dominant group. An enduring contribution of feminism to the study of science and technology is its rejection of the so-called "view from nowhere"—the misguided belief that scientific discoveries and technological advances are inherently objective and neutral (see, e.g., Gebru, 2020; Haraway, 1988; Harding, 1991; Wajcman, 1991). Historical surveys of major technological advancements, such as automobiles, medical practices, and other disciplines, reveal that because these innovations were primarily developed by the society's dominant group—white men—and reflected their supposedly universal cultural values, many have either caused harm to women or excluded them entirely (Criado Perez, 2019; Gebru, 2020). AI and other data-based algorithms seem to be going in the similar direction (O'Neil, 2016). The examples are mounting of automated hiring tools negatively biased against women, facial recognition working poorly on women of color, sexist targeted ads, deep fakes, voice assistants and robots replicating gender stereotypes. When we talk about AI-gender-based harm we, therefore, talk about two interrelated problems: (1) the inadequate representation of diverse gender groups during algorithm development, which lowers the chances of including their perspectives in the final product; and (2) the gender bias embedded in algorithmic design, which results not only from low team diversity but also dominant cultural values that suppress alternative viewpoints.

Over the last several years, there has been a surge of activity worldwide aimed at addressing the potential harms associated with AI. Governments, corporations, academics, and even religious leaders have joined the conversation, publishing a flood of frameworks, principles, and policy proposals (Jobin et al., 2019; Robert et al., 2024; Schopmans & Cupać, 2021; Tallberg et al., 2023). These documents often advocate for "responsible," "trustworthy," and "human-centered" AI, reflecting a growing consensus on the need to balance innovation with ethical considerations (Jobin et al., 2019, Smuha, 2019). The increasing awareness of AI's risks, including its gender-related implications, has also sparked calls for stronger global cooperation in AI governance, broadly understood as the process by which diverse, cross-border AI interests are harmonized in the absence of a single sovereign authority to enable collective action (Finkelstein, 1995; Weiss, 2000). While this emerging network of international AI regimes suggests a promising governing framework, significant challenges lie ahead. Interstate competition over AI dominance, the inefficiencies of global institutions, and conflicting policy such as those stemming from the populist leaders and global gender backlash continue to complicate meaningful cooperation and, importantly, the implementation of adopted measures.

Against this background, this paper seeks to map and assess how gender-based harm caused by AI is addressed in the most prominent documents of global AI governance. It examines in detail documents adopted or issued by the OECD, G20, G7, UNESCO, the EU, the Council of Europe, the European Commission's High-Level Expert Group on AI (AI HLEG), and the United Nations High-Level Committee on Programmes (HLCP). Additionally, the paper also considers global initiatives such as the Global Partnership on AI (GPAI), the World Economic Forum's AI Governance Alliance, Global AI Ethics Consortium (GAIEC), the AI4 People's Ethical Framework for a good AI society, albeit in less detail. The documents analyzed, therefore, range from binding legislation, such as the EU AI Act, to soft law instruments and non-binding guidelines. Collectively, they shape the landscape of gendered AI governance.

This paper is organized into five sections, followed by a conclusion. The first section defines the key concepts of AI and gender bias, introduces the global AI governance documents selected for analysis, and outlines the analytical approach. The second section analyzes how these documents address gender issues in the context of AI development and implementation. The third section uses these findings to identify broader trends in the integration of gender into global AI governance. The fourth discusses emerging gaps and limitations, while the fifth proposes areas for future development to strengthen gender inclusion in AI governance.

1. **Defining AI and Gender Bias: Context, Challenges, and Global Governance**

Definition of AI has been a subject of intense debate for decades and remains without a clear consensus among experts (Cardon et al., 2018). In both politics and science, the term sometimes encompasses specific systems like ChatGPT, facial recognition technology, or autonomous vehicles, while at other times, it applies to methods such as machine learning, deep learning, heuristic search, or symbolic reasoning. Further, some researchers view AI as an overarching field that includes subdomains like natural language processing, computer vision, speech recognition, and robotics. Some, however, dismiss these as overly "narrow" definitions, instead associating AI with the still-unrealized goal of machines possessing human-level or even superhuman intelligence. As a result, AI remains a fluid and ambiguous term, with any attempt to pin down a specific definition inherently open to critique (Kuhlmann et al., 2019; Schopmans, 2022). For the purposes of this paper, however, AI is conceptualized by reference to the breakthrough advancements initiated in 2012, when deep neural networks (a specific class of algorithms) started delivering significantly better results than earlier approaches across various classification tasks (Krizhevsky, Sutskever, & Hinton, 2012; LeCun, Bengio, & Hinton, 2015). These algorithms, representing a specific subset of machine learning models, achieved unprecedented improvements in predictive accuracy across a wide range of classification tasks (Cupać et al., 2024, p. 903). Unlike earlier computational approaches that relied on explicitly programmed rules, deep neural networks are trained on vast datasets comprising millions, billions, or even trillions of data points. This capacity enables them to identify complex patterns and make decisions autonomously, guided only by the system's goals. Over the course of a decade, deep learning has evolved into the foundational architecture for a range of technologies, including hiring and social media algorithms, facial recognition systems, and content moderation tools, fundamentally transforming both the operation of these systems and the ways in which users interact with them.

Despite being hailed as a transformative technology, AI is fundamentally conservative in nature, relying on historical data to learn and make decisions. Rather than generating entirely new patterns, AI systems tend to reproduce—and often entrench—existing societal structures and biases. When the underlying data reflects longstanding inequalities—whether based on gender, race, class, or other categories—AI systems inevitably learn and replicate these biases (Bengio et al., 2025). Moreover, the phenomenon of runaway feedback loops compounds this problem, as biased outputs are incorporated into new datasets, reinforcing and amplifying disparities over time (Pagan et al., 2023). Broadly speaking, these biases give rise to two types of AI harm: individual and systemic. Individual harms refer to specific, identifiable negative impacts on particular people, such as a female entrepreneur receiving lower visibility in algorithmically ranked search results. Systemic harms, by contrast, occur when AI technologies reinforce broader patterns of inequality, for instance by consistently underrepresenting women in professional recommendation algorithms (Solove, 2020; Acemoglu, 2021; Smuha, 2021; Lu, 2024; Cupać & Sienknecht, 2024). Systemic harms can also emerge from the accumulation of many individual harms, which together entrench and normalize existing structures of exclusion.

Several high-profile cases demonstrate how bias in AI systems translates into gender-based harms. In 2016, researchers working with Word2Vec—a natural language processing model trained on a large corpus of Google News texts—demonstrated how AI systems can encode and amplify gender stereotypes. The model, when tasked with completing analogies, produced outputs such as "man is to computer programmer as woman is to homemaker" (Bolukbasi et al., 2016), embedding patriarchal assumptions at a systemic level. In 2018, Amazon was forced to abandon an experimental AI hiring tool after it was found to disadvantage women applicants. Trained on historical hiring data dominated by men, the system penalized resumes that included references to women's organizations, inflicting harm on individual applicants while creating conditions for perpetuating systemic gender imbalances (Dastin, 2016). That same year, Joy Buolamwini and Timnit Gebru exposed significant racial and gender disparities in commercial facial recognition software, showing that systems misclassified dark-skinned women far more often than light-skinned men (Buolamwini & Gebru, 2018). Their work also drew important attention to *intersectionality*—a concept introduced by Kimberlé Crenshaw—to show how overlapping forms of discrimination, such as those based on race, class, and gender, create unique experiences of marginalization (Crenshaw, 1989).

AI gender harm cannot, however, be attributed solely to biased datasets. Despite their apparent autonomy, these systems are fundamentally shaped by the decisions and intentions of their human creators. The biases of individual developers, dominant cultural values, and the structural inequalities within the AI industry, influence choices around data selection, labeling, and usage, ultimately leading to the creation of discriminatory AI systems. In other words, the lack of gender diversity in AI engineering and leadership is another reason why we see the perpetuation of patriarchal patterns within these digital systems (Fazelpour & De-Arteaga, 2022; Li, 2020; D'Ignazio & Klein 2020). Globally, women make up just 22% of the AI workforce, with their presence dropping further at senior levels, where they hold under 14% of executive positions in the field. The situation is similarly troubling in the European Union, where the gender gap in AI talent is even wider than in the general labor market. Germany and Sweden, for example, report some of the lowest proportions of women in AI roles across the EU, at 20.3% and 22.4% respectively (Pal et al., 2024). In response to this persistent exclusion, feminist activists and scholars have amplified calls for greater inclusivity under the slogan "nothing about us without us"—a phrase originating in the disability rights movement—emphasizing that no software, decision, or policy should be developed without the active participation of those directly affected (Costanza-Chock, 2023; Charlton, 1998).

Against this background, this paper examines how global AI governance documents and initiatives have addressed concerns related to gender bias in AI systems and the persistent underrepresentation of women in AI development and leadership. Over the past several years, the number and variety of these documents and initiatives have expanded significantly, encompassing a range of instruments. These include binding laws, such as the EU AI Act; binding treaties, such as the Council of Europe's Framework Convention on Artificial Intelligence and Human Rights, Democracy, and the Rule of Law; and soft international law instruments, including the OECD Council's Recommendation on Artificial Intelligence, the G20 Ministerial Statement on Trade and Digital Economy, UNESCO's Recommendation on the Ethics of Artificial Intelligence, and the G7 Hiroshima AI Process. In addition, a variety of guidelines, frameworks, ethical standards, and both state and non-state initiatives have emerged, such as the Ethical Guidelines for Trustworthy AI by the European Commission's High-Level Expert Group on AI (AI HLEG), the Principles for the Ethical Use of Artificial Intelligence in the United Nations System by the United Nations High-Level Committee on Programmes (HLCP), the Global Partnership on AI (GPAI), the World Economic Forum's AI Governance Alliance, Global AI Ethics Consortium, and the AI4 People's Ethical Framework for a good AI society.

A common feature across many—though not all—of these documents is their reliance on a risk-based model of AI governance, which has recently emerged as a dominant approach. Risk-based frameworks tend to focus on identifying and mitigating discrete, usually quantifiable types of harms. While this model is valuable for addressing certain risks, it often reflects an individualized understanding of harm and can overlook broader systemic patterns of discrimination (Smuha, 2021; Lu, 2022). Although this paper's analysis reflects the distinction between individual and systemic harms and how risk-based approaches relate to them, its primary focus lies elsewhere. The aim is to offer a broader snapshot of how gender concerns are being addressed in global AI governance. A deeper exploration of the type and nature of regulatory frameworks would require a level of analysis that is beyond the scope of this paper, and thus best left for further research.

To carry out the analysis, I conducted a systematic qualitative review of each document, scanning for explicit references to gender in relation to AI system development and implementation. I also examined broader normative concepts—such as non-discrimination, equality, fairness, and individual autonomy—that, while not always directly linked to gender, have clear implications for gender-sensitive AI governance. This dual approach made it possible to capture both explicit and implicit engagements with gender-related concerns. Based on this review, I identified five emerging gender-related trends in global AI governance, along with notable gaps that warrant further attention as the field evolves. While the primary focus is on binding laws, treaties, and soft law instruments, relevant AI guidelines, frameworks, ethical standards, and initiatives are also included to provide a more comprehensive understanding of the place of gender in global AI governance. The final section outlines potential paths forward for the integration of gender perspectives in future AI governance efforts.

## 2. How Do Global AI Governance Documents Address Gender Issues?

### 2.1. The OECD Council's Recommendation on Artificial Intelligence

Adopted in 2019 and amended in 2024, the OECD Council's Recommendations on Artificial Intelligence represent the first intergovernmental standard on AI aimed at shaping policy among its 38 member states and beyond. The document is renowned for advancing five principles designed to foster innovation and trust in AI by promoting the "responsible stewardship of trustworthy AI while ensuring respect for human rights and democratic values" (OECD, 2024). The principles are as follows: (1) inclusive growth, sustainable development and well-being; (2) respect for the rule of law, human rights and democratic values, including fairness and privacy; (3) transparency and explainability; (4) robustness, security and safety; and (5) accountability (OECD, 2024).

The Recommendations mention gender only once, as part of the first principle—inclusive growth, sustainable development, and well-being—whereby they ask stakeholders to "advance inclusion of underrepresented populations," thus "reducing economic, social, gender, and other inequalities" (OECD, 2024). However, gender harm and inclusion can be understood as implicitly covered by the other principles due to their general nature. For example, the second principle demands that AI actors respect the rule of law, human rights, and democratic and human-centric values, principles inextricably tied to "non-discrimination, equality, freedom, dignity, autonomy of individuals, privacy and data protection, diversity, fairness, social justice, and internationally recognized labor rights" (OECD, 2024). Each of these principles can be invoked to prevent or remedy gender harm produced by AI systems. Similarly, the third principle—transparency and explainability—might aid in identifying and mitigating gender harm embedded within algorithms and data sets. In short, the Recommendation of the OECD Council on Artificial Intelligence is a brief and general document which, rather than paying focused attention to bias—including gender bias—aims to offer a broad set of principles under which gender bias and other types of gender harms could be identified and mitigated in different contexts.

**2.2. G20 Ministerial Statement on Trade and Digital Economy**

In its 2019 document titled "G20 Ministerial Statement on Trade and Digital Economy," the G20 participating states included an annex titled G20 AI Principles, in which the document largely draws—or more precisely, copies five of—the OECD principles (G20, 2019). Accordingly, gender is not given much explicit attention. It is mentioned once in the first principle—inclusive growth, sustainable development, and well-being—repeating what is said in OECD Recommendations about the need to advance the "inclusion of underrepresented populations" thus "reducing economic, social, gender, and other inequalities." As with the OECD's Recommendation, it is expected that the other four principles cover gender inclusion and harm implicitly through broad social principles such as non-discrimination, transparency, and accountability. However, the body of the G20 Ministerial Statement goes a step further, with an eye on the digital gender divide and the need to overcome it with more urgency. The Statement emphasizes gender inclusivity in the digital economy by demanding equitable access to technology for women and girls. This focus is reflected in the G20's support for initiatives such as EQUALS and G20 #eSkills4girls, which aims to enhance women's participation in the digital economy by providing digital skills training and access to technology (G20, 2019, p.6) The G20 Statement is therefore chiefly preoccupied with developing an economic case for gender equality in AI, insisting that promoting gender equality in digital skills and employment strengthens economic growth, innovation, and social inclusion.

## 2.3. UNESCO Recommendations on the Ethics of Artificial Intelligence

The UNESCO Recommendations on the Ethics of Artificial Intelligence, adopted in 2021, represent the first *global standard* on AI ethics, applicable to all 194 UNESCO member states (UNESCO, 2021). In addition to outlining broad ethical principles, the Recommendations provide an expansive discussion of policy actions designed to help policymakers translate these principles and core values into practice. These actions address issues such as data governance, environmental sustainability, education and research, health and well-being, and, of particular interest here, gender (UNESCO, 2021, p. 32). The Recommendations not only explicitly identify gender as an area of concern in AI development and implementation but also dedicate a comprehensive and robust framework to addressing it. This framework recognizes both the need to empower marginalized groups to participate in AI development and decision-making and the importance of eliminating gender bias and other harms from AI systems and the data on which they are based. Accordingly, the UNESCO Recommendations emphasize that AI should not exacerbate the existing gender divide, such as the wage gap or unequal representation in certain professions; it should also not perpetuate gender stereotypes or discriminatory biases, nor should it violate the human rights and fundamental freedoms of girls and women in any way (UNESCO, 2021, p. 32, Articles 89, 90, 87). Instead, the Recommendations assert that AI's potential should be fully harnessed to contribute to gender equality (UNESCO, 2021, p. 32, Article 87). To this end, member states are expected to empower women and girls to participate in the digital economy and AI development, encourage their entrepreneurship and involvement in all stages of the AI system life cycle, and promote gender diversity in research, academia, and industry (UNESCO, 2021, p. 32, Article 88, 91, 92). Unlike the OECD's Recommendation and the G20's Statement, UNESCO's Recommendations place gender at the forefront of AI policy. While not perfect, these recommendations, nonetheless, leave less room for interpretation regarding the gender's place in development and implementation of AI systems.

## 2.4. G7 Hiroshima AI Process

The Hiroshima AI Process, initiated by the G7 in May 2023, aims to establish a coordinated and comprehensive global framework for promoting and developing safe, secure, and trustworthy AI, particularly focusing on advanced AI models like generative AI. The process consists of four pillars: analysis of priority risks, challenges and opportunities of generative AI; the International

Guiding Principles for all AI actors in the AI ecosystem; the International Code of Conduct for Organizations Developing Advanced AI Systems; and project based cooperation in support of the development of responsible AI tools and best practices. The *Guiding Principles* and the *Code of Conduct*, the two documents adopted in the Hiroshima process, provide insight into how the G7 views gender in relation to AI development (G7, 2023a; G7, 2023b). Notably, neither document explicitly mentions gender. However, gender can be considered implicitly addressed through the broader risk-based approach to AI adopted by the participating states, which ties risk to safety, security, human rights, and broader principles such as democracy, the rule of law, diversity, fairness, and non-discrimination. In this sense, the twelve guiding principles—including obligations to identify, evaluate, and mitigate risks and vulnerabilities, report incidents caused by AI systems, ensure content authentication, maintain data quality, and protect personal data—can be extended to address harms AI may cause with respect to gender (G7, 2023a, p. 2-5). The same applies to the *Code of Conduct*. While it professes a general commitment to fairness, non-discrimination, the prevention of harmful bias, and promotes human-centric AI design, it does not specifically address gender—it needs to be inferred (G7, 2023b). Moreover, the *Code of Conduct does* not explicitly call for diversity within AI developer teams.

## 2.5. The EU Artificial Intelligence Act

The EU AI Act, which came into effect in August 2024, establishes a unified regulatory and legal framework for artificial intelligence within the European Union (EU, 2024). In comparison to other documents, this legal instrument pays significant attention to gender. In Recital 27, it thus makes a general point by explicitly stating that "[d]iversity, non-discrimination and fairness means that AI systems are developed and used in a way that includes diverse actors and promotes equal access, *gender equality* and cultural diversity, while avoiding discriminatory impacts and *unfair biases*" (EU, 2024, Recital 27, emphasis added). In Article 68 and Recitals 165, respectively, the Act also calls for diversity, including "gender balance," to be fostered in teams developing AI and scientific panels of independent experts (EU, 2024, Article 38, Recital 165).

Crucially, the Act recognizes the importance of gender in all three categories of its risk-based approach to regulating AI: unacceptable, high and low or minimal risk. Risks deemed unacceptable—and therefore requiring the prohibition of related AI systems —include systems that deploy subliminal techniques; exploit vulnerabilities such as age, disability, or socio-economic situation; evaluate and classify individuals based on their social behavior or personal characteristics; certain facial recognition systems; systems that infer emotions within workplace and educational institutions; biometric categorization systems; and real-time remote biometric identification systems (EU, 2024, Article 5). In each of these categories, gender can serve as a data point used toward specific malicious outcomes, thus becoming a gateway to harm by reinforcing stereotypes, enabling discrimination, or amplifying biases. Similarly, most areas identified in the Act as high-risk AI—particularly those related to education, employment, public services, law enforcement, migration, and the administration of justice—are recognized as domains where the use of AI may give rise to new forms of gender discrimination or perpetuate historical patterns of inequality (EU, 2024, Annex I and Annex II). This risk, among other factors, is precisely what classifies AI deployed in these areas as high-risk. Finally, even in the area of minimal AI risk, the EU AI Act asks developers and deployers to assess and prevent the negative impact of AI systems on gender equality (EU, 2024, Article 95). In summary, while the effectiveness of these provisions—implemented through a robust risk management system that includes conformity assessments, fundamental rights impact assessments for high-risk AI, regulatory sandboxes, and post-market deployment monitoring—remains to be evaluated, the EU AI Act takes a comprehensive approach to addressing gender and AI; it not only recognizes the nature of these risks but also acknowledges their potential to permeate various social domains.

### 2.6. The Council of Europe Framework Convention on Artificial Intelligence and Human Rights, Democracy and The Rule of Law

The Council of Europe Framework Convention on Artificial Intelligence, Human Rights, Democracy, and the Rule of Law stands as the first legally binding international treaty on AI (Council of Europe, 2024). Opened for signature on September 5, 2024, the treaty aims, as its name suggests, to ensure that all activities across the AI lifecycle adhere to the principles of human rights, democracy, and the rule of law, while fostering technological progress and innovation. The treaty establishes *principles* related to activities within the lifecycle of AI systems, proposes *remedies* and procedural safeguards, and framework for *assessment and mitigation of AI risks* and adverse impacts (Council of Europe, 2024).

Gender is explicitly tackled only in the principles section under the principle titled "equality and discrimination." While all other principles cited in the treaty—human dignity and autonomy, transparency and oversight, accountability and responsibility, privacy and personal data protection, reliability, and safe innovation—can be evoked to uphold gender standards in AI development and deployment, the principle titled "equality and discrimination" explicitly states that "[e]ach Party shall adopt or maintain measures with a view to ensuring that activities within the lifecycle of artificial intelligence systems respect equality, including *gender equality*, and the prohibition of discrimination" (Council of Europe, 2024, Article 10, emphasis added). Gender is not explicitly addressed in the treaty's remedies section. However, since it makes harm to any type of human rights eligible for redress, it can be interpreted to also encompass gender-based AI harm. This approach, therefore, implicitly includes gender-sensitive protections by emphasizing the availability of legal recourse and procedural transparency to ensure that AI system decisions do not unfairly disadvantage individuals based on gender. The same applies to the treaty's section on the assessment and mitigation of AI risk. Gender is not explicitly mentioned here either, but it falls within the scope of the obligations requiring risk assessments to consider the impact of AI on various vulnerable groups. Particular emphasis is placed on addressing discriminatory or biased outcomes that may arise, including those related to gender contexts.

## 2.7. Guiding Frameworks, Ethical Principles, and Initiatives for AI Development and Deployment

Besides binding laws on AI, such as the EU AI Act, and soft law instruments like the UNESCO Recommendations, there exist numerous guidelines, frameworks, ethical principles, and initiatives that aim to shape how AI is conceptualized and developed. Notable among these are the *Ethics Guidelines for Trustworthy AI* adopted in 2019 by the European Commission's High-Level Expert Group on AI (HLEG on AI) and the *Principles for the Ethical Use of Artificial Intelligence in the United Nations System* adopted in 2021 by the United Nations High-level Committee on Programmes (HLCP) (HLEG on AI, 2019; HLCP, 2021). Both documents integrate gender considerations as part of their broader frameworks for AI fairness, non-discrimination, and inclusivity, but they approach the issue with distinct emphases and operational strategies.

While both frameworks advocate for gender equality in AI systems, the UN Principles place a more explicit emphasis on promoting gender equality as a core objective, integrating it directly into their ethical use recommendations (HLCP, 2021, p. 6). This approach includes meaningful consultations with affected gender groups, thereby emphasizing a participatory and inclusive framework. In contrast, the European Commission addresses gender inclusivity within broader principles of fairness and non-discrimination, embedding it as one aspect among various dimensions of inclusivity rather than isolating it as a standalone priority. The Ethics Guidelines for Trustworthy AI stress the avoidance of unfairly biased outputs and the support of diverse populations, including gender-specific groups, through stakeholder participation (HLEG on AI, 2019, p. 12 and 19). They also advocate for inclusivity via the application of Universal Design principles to accommodate diverse characteristics, including gender, and emphasize accountability measures, such as continuous evaluation and stakeholder involvement to address gender biases (HLEG on AI, 2019, p. 19 and 24). Similarly, the UN Principles underscore fairness by advocating for the prevention of bias and stigmatization, with particular attention to ensuring the equal and just distribution of AI's benefits across genders (HLCP, 2021, p. 5). They also adopt an interdisciplinary and participatory approach to AI design, linking inclusion explicitly with gender equality and stakeholder consultation (HLCP, 2021, p. 6). Furthermore, the UN framework calls for governance structures and audits that integrate gender considerations as part of their accountability mechanisms (HLCP, 2021, p. 6). Despite differences, both frameworks underscore the importance of participatory processes and continuous oversight to achieve equitable outcomes in AI systems.

Several global initiatives address gender in AI governance, focusing on inclusivity, bias mitigation, and equitable representation. *The Global Partnership on AI (GPAI)*, launched in 2020 by OECD member states and others, promotes gender equality by encouraging diverse participation in AI development and combating algorithmic biases. *The World Economic Forum's AI Governance Alliance*, introduced in 2023, emphasizes ethical frameworks that integrate gender perspectives to prevent AI systems from perpetuating discrimination. *The Global AI Ethics Consortium (GAIEC)*, initiated in 2018 by a coalition of academic and policy organizations, advocates for ethical AI through frameworks that include gender as a critical axis of fairness and equity. Similarly, *AI4People's Ethical Framework for a Good AI Society*, launched in 2018 by the Atomium-EISMD initiative, incorporates gender as a key ethical consideration, urging stakeholders to address structural inequalities through AI design and deployment. Although not perfectly, together, each of these initiatives highlight the importance of embedding gender-sensitive approaches within AI development and deployment.

## 3. Identifying Gender Trends in Global AI Governance

### 3.1. Integration with Broader Human Rights and Ethical Frameworks

One of the first trends that can be observed is that gender equality in AI is increasingly integrated within broader and comprehensive human rights and ethical frameworks and principles, such as justice, fairness, and non-discrimination. We can see this in the OECD and G20 AI Principles, which emphasize inclusive growth, sustainable development, rule of law, human rights, democratic values, fairness, privacy, and so on; the Council of Europe's Framework Convention on Artificial Intelligence which highlights human dignity, autonomy, equality, and non-discrimination; and the EU AI Act which underscores that all AI development and deployment should happen within the scope of the EU's fundamental rights. This trend demonstrates that gender equality, particularly as it pertains to AI development and deployment, is not an isolated issue but a critical aspect of broader human rights and ethical frameworks, as well as comprehensive ethical guidelines on AI.

The integration of gender considerations within broader human rights frameworks enhances the moral imperative to address gender biases in AI. For example, this approach aligns AI governance with established international commitments, such as the Sustainable Development Goals (SDGs), specifically Goal 5: *Achieve gender equality and empower all women and girls*. Another notable framework that serves this purpose is the UN Guiding Principles on Business and Human Rights which provides a framework for corporations to respect human rights, including gender equality. These principles have been adapted to AI through initiatives like the Business for Social Responsibility (BSR) report on "Artificial Intelligence: A Rights-Based Blueprint for Business," which guides companies in implementing AI responsibly with respect to human rights, including gender rights.

### 3.2. Increasing Explicitness in Addressing Gender

However, the trend towards embedding global AI governance within broader human rights and ethical frameworks and principles intersects with another trend: a shift from an implicit to an explicit treatment of gender issues. This change is rooted not only in the policy culture, style, and orientation of documents produced by international organizations but also reflects a growing body of evidence highlighting how AI systems can inadvertently perpetuate gender biases present in society. As shown previously, as various types of AI systems proliferate, it is becoming increasingly clear that natural language processing models trained on biased datasets may generate sexist language or reinforce gender stereotypes. High-profile incidents, such as Amazon's discontinued AI recruiting tool that favored male candidates due to biased training data, have also underscored the urgent need for explicit gender considerations in AI development and governance. These events have alerted policymakers, technologists, and civil society to increase pressure and seek greater integration of gender perspectives into AI policies and frameworks.

Accordingly, early documents on AI governance—namely the OECD and G20 AI Principles—deal with gender matters in AI with little or no direct acknowledgment, addressing them instead implicitly through, as noted earlier, broader principles such as fairness, non-discrimination, and inclusive growth. More recent documents, such as the AI4People Ethical Framework, the UN HLCP's Principles for the Ethical Use of Artificial Intelligence in the United Nations System, the UNESCO Recommendation on the Ethics of Artificial Intelligence, the EU AI Act, and the Council of Europe's Framework Convention on Artificial Intelligence and Human Rights, Democracy, and the Rule of Law, tackle gender in relation to AI more explicitly, both by embedding it in broader principles and recognizing its standalone significance. Most of these texts also provide detailed guidelines aimed at promoting gender equality and mitigating biases in AI systems. This shift can be read as an indication of a growing awareness of the pervasive issue of gender bias in AI and a clear movement from implicit to explicit inclusion of gender-related considerations in global AI governance.

### 3.3. Enhanced Focus on Inclusivity, Diversity, and Empowerment

Another notable trend in AI governance as it relates to gender is the emphasis on inclusivity and diversity as foundational principles. This reflects an acknowledgment that a shift is needed from merely preventing discrimination to actively fostering gender diversity and inclusivity in both AI education, development, and governance roles, with the goal of ensuring that diverse voices are represented in shaping AI systems. As discussed earlier, documents such as the OECD AI Principles, the UNESCO Recommendation on the Ethics of Artificial Intelligence, and the EU AI Act, each have, more or less explicit provisions, which put emphasis on the female participation in AI development and research, with UNESCO also putting particular emphasis on empowering girls to be a part of digital economy and AI development. Other initiatives and documents such as the Global Partnership on AI (GPAI) and the World Economic Forum's (WEF) Global AI Ethics and Governance Framework also advocate for integrating gender diversity within AI development teams to incorporate a range of perspectives. Complementing this, initiatives like AI4ALL's "Gender and AI" and Women in AI (WAI) under the Global AI Ethics Consortium (GAIEC) focus on increasing gender representation in the AI field.

Importantly, inclusivity and diversity in AI development and research are increasingly being operationalized through concrete initiatives and policies. Companies like Microsoft and Google have implemented diversity hiring goals and inclusive workplace policies to foster a more diverse AI workforce (Google, 2022; Microsoft, 2024). Research institutions are launching programs to support underrepresented groups in AI. For example, the Alan Turing Institute in the UK has established fellowships and scholarships aimed at increasing diversity in AI research.[1] These programs provide financial support, mentorship, and networking opportunities, helping to break down barriers to entry and advancement. At the policy level, the EU's Horizon Europe research and innovation framework includes provisions for gender equality.[2] Projects funded under this program are required to integrate gender dimensions into their research and innovation content, promoting the development of AI technologies that consider the needs and experiences of all genders. Educational institutions too are implementing programs to attract and retain women in AI-related fields. For example, Carnegie Mellon University's Women in AI initiative provides scholarships, mentorship, and research opportunities specifically for female students. Finaly, training programs and workshops, such as those offered by the AI Ethics Lab, help developers understand how biases, including gender bias, can enter AI systems and how to prevent them.

---

[1] See: https://www.turing.ac.uk/people/fellows/turing-ai-fellows
[2] See: https://research-and-innovation.ec.europa.eu/strategy/strategy-research-and-innovation/democracy-and-rights/gender-equality-research-and-innovation_en?utm_source=chatgpt.com

### 3.4. Emphasis on Practical Implementation and Accountability

Documents and initiatives within global AI governance are increasingly emphasizing practical measures to address gender biases. This represents a shift from merely setting principles to ensuring practical implementation and sustained accountability. The EU AI Act, being a legally binding document, has advanced the furthest in this regard. Although it does not single out gender bias specifically, it addresses the issue within the broader context of AI-related harms. The Act establishes multiple transparency and risk management mechanisms, including: visibly labeling AI systems that interact with humans; placing transparency labels on AI-generated content such as deepfakes and synthetic text; requiring developers to be transparent about the data used to train AI systems; conducting conformity assessments and fundamental rights impact assessments; and implementing regulatory sandboxes, human oversight, and post-market surveillance (EU, 2024).

Here it should be noted that global AI governance in terms of the implementation of principles is supplemented by domestic AI policies. Accordingly, to translate principles into practice, several governments are developing tools and regulatory mechanisms. For example, the UK's Information Commissioner's Office (ICO) has published in 2020 guidance on *Explaining decisions made with AI*, which includes considerations for fairness and avoiding discrimination, including gender bias. This guidance helps organizations understand their legal obligations under the UK's Equality Act and Data Protection Act when deploying AI systems. In the United States, in 2023, the National Institute of Standards and Technology (NIST) issued a voluntary *AI Risk Management Framework* aimed to provide organizations with a framework and strategies to identify, assess, and manage risks, including those related to gender biases.

### 3.5. Collaboration and Global Standards

AI governance is increasingly shaped by international cooperation and standardization. Organizations like GPAI, UNESCO, and WEF stress the need for harmonized gender-inclusive AI standards to address gender biases globally. Efforts by groups such as the Global AI Ethics Consortium (GAIEC) seek to create standardized frameworks that promote consistency and shared accountability across nations. This trend demonstrates a growing emphasis on global collaboration to ensure that gender inclusivity is a universal priority in AI governance. International organizations are also spearheading efforts to develop global standards for AI governance that include gender considerations. The OECD's AI Policy Observatory serves as a platform for sharing policy insights and promoting international cooperation. It provides resources and tools to help governments implement the OECD AI Principles, which, while not gender-specific, promote inclusive growth and fairness. Standards bodies like the IEEE are developing standards that incorporate ethical considerations into AI design. The IEEE's "Ethically Aligned Design" initiative includes recommendations for ensuring AI respects human rights and values, including gender equality. These standards provide technical guidelines for developers to implement ethical principles in practice.

While these trends reflect promising developments in global AI governance, it remains uncertain whether they will endure. Rising backlash against the Liberal International Order, growing domestic and institutional anti-feminist movements, and deregulatory pushes prioritizing technological freedom over oversight all threaten recent gains (see, e.g.: Alter & Zürn, 2020a, 2020b; Kuhar & Paternotte, 2017; Cupać & Ebetürk, 2021, 2020). The Trump administration's efforts to remove references to gender and fairness from AI governance illustrate how quickly progress can be reversed (Heaven, 2024). At a time when obstacles to meaningful progress are only increasing, keeping gender equality at the center of global AI governance will, therefore, require sustained advocacy, strong political will, and international cooperation wherever possible.

### 4. Gaps in Global AI Governance

Although some trends in global AI governance signal positive developments in addressing gender-related AI harm, significant problems remain. First, while recently adopted governance documents more explicitly address gender and efforts to standardize AI governance globally are ongoing, global AI governance still suffers from the inconsistent inclusion of gender across different governing frameworks. This uneven treatment creates a patchwork of regulations, making it unlikely to uniformly combat AI-produced gender bias across various international organizations, states, and AI developers. Additionally, this patchwork governing framework is complicated by cultural and regional variations in concepts of gender and gender equality. In traditional societies with rigid gender roles or those experiencing gender backlash due to rising right-wing, conservative, populist, and religious forces, promoting gender diversity in AI may face significant resistance. Consequently, policies that are welcomed and effective in one region may not be applicable or accepted in another.

Second, a related issue is the infrequent recognition of the importance of intersectionality—i.e., how overlapping social identities such as race, gender, class, and sexuality interact to create unique experiences of discrimination and privilege—in these global AI governance documents. While some documents, such as UNESCO's Recommendation on the Ethics of AI and the AI4People Ethical Framework, engage with intersectionality and its compounding impact on gender harm produced by AI systems, the majority of governing documents do not venture in this direction. Yet, without the recognition of intersectionality in the treatment of gender, it will not be possible to speak about a robust approach to addressing AI-related gender harm.

Third, general statements about promoting fairness and non-discrimination, including those related to gender issues, which are frequent in global AI governance documents are unlikely to translate into practical measures without detailed instructions. This is exacerbated by the generally weak enforcement mechanisms envisioned in global AI governance. Even where policies exist, enforcement may be lacking. Voluntary guidelines and self-regulation are often insufficient to ensure compliance, especially when addressing deeply ingrained biases. Without legal mandates or the threat of penalties, organizations may not prioritize gender considerations, particularly if they perceive them as conflicting with other business objectives. Further compounding the issue is that regulators still lack the technical expertise or resources to investigate complex AI systems, and affected individuals may not be aware of their rights or how to assert them.

Fourth, despite some positive trends and multiple initiatives, underrepresentation in leadership positions perpetuates a cycle in which the perspectives of women and gender minorities are not adequately reflected in AI governance. As indicated earlier, this lack of diversity at the decision-making level is likely to result in blind spots and a failure to prioritize gender issues. Furthermore, efforts to promote diversity, while laudable, often focus on entry-level positions; without pathways to advancement, these initiatives may have limited impact. Structural barriers such as unconscious bias, lack of mentorship opportunities, and non-inclusive workplace cultures hinder the progression of underrepresented groups to leadership roles.

Finally, rapid technological advancements in AI introduce new dimensions of gender bias. Emerging technologies, such as deepfakes, autonomous systems, and AI-driven surveillance, can have unforeseen impacts on gender equality. For instance, deepfake technology has been misused to create non-consensual explicit content targeting women, posing significant privacy and safety concerns. Regulatory frameworks often lag behind technological advancements, leaving gaps where new forms of gender bias can emerge unchecked.

## 5. Future Directions for Enhancing Gender Protection in AI Governance

Analyzing current trends in global AI governance shows that while there has been progress, significant gaps remain when it comes to integrating gender perspectives. The full harmonization of AI governance globally—especially regarding gender issues—is probably out of reach. However, it is still crucial to find a workable balance between promoting universal principles of gender equality and respecting local cultural contexts. Three main areas thus stand out for strengthening gender-sensitive global AI governance.

**First, building explicit gender-responsive governance frameworks**. IOs such as UNESCO, the UN Human Rights Council, and the OECD can lead this process by developing detailed non-binding instruments—such as model guidelines, recommendations, and frameworks—that member states can incorporate into national legislation. These frameworks should help institutionalize gender and intersectional perspectives at all stages of AI development and deployment. For example, *intersectional impact assessments* could be made mandatory to identify layered biases, while specific methodologies, including *algorithmic audits* and *fairness testing*, could be recommended to detect and correct gender imbalances in AI systems. *Ethical data governance protocols* that promote privacy while enabling the identification of gender-related biases should also be established. To reinforce these standards, IOs could advocate for independent regulatory bodies at the national or regional level, capable of monitoring compliance, investigating complaints, and enforcing gender equality provisions through sanctions or other corrective measures.

**Second, sustainable integration of gender perspectives also demands more profound shifts in leadership, research, and evaluation practices.** IOs, in collaboration with regional bodies, academic institutions, and civil society organizations, could work to diversify leadership within AI sectors by promoting training programs, mentorship initiatives, and inclusive recruitment practices. Procurement standards and certification schemes could be used to reward organizations that meet diversity targets. At the same time, measuring progress matters. Standardizing indicators—such as gender representation in AI teams, the diversity of datasets, and the fairness of AI outcomes—can make it easier to track whether commitments are turning into real change. Beyond leadership and metrics, supporting interdisciplinary research is key. Funding initiatives that bring together technical experts and social scientists, including those from gender studies, could help deepen understanding of how AI systems interact with existing inequalities—and how they might do better.

**Third, governance approaches must stay flexible and future-oriented to keep up with rapid technological change.** As AI technologies evolve rapidly, IOs have a critical role in coordinating foresight activities, such as horizon scanning and scenario planning, that specifically consider gender implications. Establishing regulatory sandboxes where new AI systems can be tested for gender biases before full deployment would allow policymakers to proactively identify risks early and mitigate harms. At the same time, broader public engagement is essential to building support for gender-equitable AI governance. IOs could launch global public awareness campaigns to inform individuals about their rights concerning AI technologies and provide accessible educational resources tailored to policymakers, educators, and diverse communities. Working closely with local stakeholders—especially feminist networks and marginalized communities—ensures that governance frameworks are not technocratic but responsive to the lived experiences of those most affected.

Of course, implementing these suggestions does not guarantee the full mitigation of individual or systemic gender harms linked to AI. The challenges are not only technical but, as mentioned above, also deeply political. In a context where anti-gender backlash is rising across international organizations such as the UN and the EU, as well as in many states, resistance to developing robust gender protections in AI governance is likely to grow (Kuhar & Paternotte, 2017; Cupać & Ebetürk, 2021, 2020). Nevertheless, these obstacles should not deter efforts to deepen our understanding of how gendered harms are produced and sustained through AI systems, nor discourage the development—however preliminary or aspirational—of regulatory approaches that could help mitigate such harms. Even if progress is incremental, mapping out possibilities for more just and equitable governance remains an essential task.

**Conclusion**

The global AI governance regime is still in its infancy. Its goals are highly complex, as it must address the individual and systemic harms AI can produce, harness AI for economic development, and navigate the power competition among the world's major superpowers. Within this framework, the issue of gender and AI governance is paramount. Failing to address gender risks perpetuating entrenched discriminatory and patriarchal patterns, while also creating new forms of bias and inequality across various areas of life, from the job market to medical practice. Against this background, this article examines key global AI governance documents and initiatives, offering an early snapshot of how they address gender issues, identifying prevailing trends and omissions, and proposing further actions to prevent AI from reproducing and amplifying harm and injustice. Effective AI governance that is sensitive to gender issues must begin with a robust understanding of how gender bias and discrimination emerge during both the development and implementation of AI systems. Neglecting this risks not only perpetuating gender bias but also leading to tokenism or ethics washing, where performative feminist gestures serve to mask injustice rather than address it, as was the case with Google's treatment of AI ethics researcher Timnit Gebru, where public commitments to diversity and fairness were undermined by internal resistance to substantive change (Hao, 2020).

Therefore, a glaring omission in most examined documents is the failure to account for how intersecting identities—such as race, class, and gender—shape people's experiences of oppression and discrimination. Although frameworks like UNESCO's Recommendations acknowledge intersectional impacts, most still overlook this critical dimension, leaving them poorly equipped to address the compounded harms AI systems can produce. Yet intersectionality is only part of the problem. Global AI governance remains fragmented, with most efforts relying on soft law rather than enforceable legal standards. Apart from a few exceptions, such as the EU AI Act and the Council of Europe's Framework Convention, binding rules and strong enforcement mechanisms are largely missing. As a result, gender considerations are often sidelined when they clash with commercial interests. Finally, the widespread reliance on risk-based governance models in these documents tends to frame AI harms in narrower, individualistic terms, making it even harder to address the deeper structural inequalities that technologies can reinforce.

Despite these shortcomings, there are positive trends to build on. Gender is increasingly addressed in global AI documents and integrated into broader human rights frameworks emphasizing diversity and empowerment. Efforts to improve accountability and standard-setting are also gaining momentum. Yet sustaining this progress demands broad action. Feminist voices must grow louder, especially amid rising anti-gender backlash, and governments must adopt frameworks that take intersectionality seriously, strengthen enforcement, and adapt to technological change. Future research will be critical in examining how individual and systemic AI gender harms are interrelated, and in supporting, together with decision-makers, the development of alternatives to risk-based governance models—such as rights-based approaches, participatory frameworks, and structural interventions that address the roots of inequality in AI systems.